"Post-Westgate SWAT : C4ISTAR Architectural Framework for Autonomous Network Integrated Multifaceted Warfighting Solutions

Version 1.0" : A Peer-Reviewed Monograph [31]

by

Nyagudi Musandu Nyagudi - Security Analyst

Nairobi, Kenya – EAST AFRICA

**Reader :**

David K. Ngondi, works at the Directorate of Police Reforms of the Kenya Police Service and has several decades of experience in a broad-spectrum of security and paramilitary operations. He possess a Master of Arts degree in Security Studies from the University of Hull in the UK.

Released on 7[th]/11/13

**Keywords:** Informatics, Information Systems, Computing, Signal Processing, Control, Forensics, Military, War, Robotics, Ethics, Automation, Autonomous, Homing, C4ISTAR, C4ISR, Battle-space Digitization, Security Studies, Future Concepts, Innovation, UUV, UAV, UGV, Systems Analysis and Design, Systems Architecture, Robots, Drones, Counter-terrorism, Urban Warfare, Room-to-Room combat,

# Contents

## List of Figures

## List of Abbreviations

**C4ISTAR** – Command, Control, Communications, Computers (Computing), Intelligence, Surveillance, Target Acquisition and Reconnaissance

**KRC** - "Killer Robots Campaign" - the Campaign Against Killer Robots

**N** – natural numbers / countable numbers

**NP** – non-deterministic polynomial time

**OODA loop** – (observe → orient → decide → act) loop

**#P** – sharp "P"

**R** – real numbers

**SAD** – Systems Analysis and Design

**SWAT** – Special Weapons and Tactics

**UAV** – Unmanned Aerial Vehicle

**UGV** – Unmanned Ground Vehicle

**UUV** – Unmanned Underwater Vehicle

| a | - modulus a

>> - much greater than

Abstract

**"Post-Westgate SWAT : C4ISTAR Architectural Framework for Autonomous Network Integrated Multifaceted Warfighting Solutions**
**Version 1.0" : A Peer-Reviewed Monograph**
by
Nyagudi Musandu Nyagudi – Security Analyst

Nations are today challenged with multiple constraints such as declining population and financial austerity, these inevitably reduce military/security forces preparedness. Faced with well resourced adversaries or those of the asymmetric type, only a Nation that arms itself "intelligently" and fights "smart" attains advantages in the world's ever more complex and restrictive battle-spaces. Police SWAT teams and Military Special Forces face mounting pressure and challenges from adversaries that can only be resolved by way of ever more sophisticated inputs into tactical operations

Lethal Autonomy provides constrained military/security forces with a viable option, but only if implementation has got proper empirically supported foundations. Autonomous weapon systems can be designed and developed to conduct ground, air and naval operations. This monograph offers some insights into the challenges of developing legal, reliable and ethical forms of autonomous weapons, that address the gap between Police/Law Enforcement and Military operations that is growing exponentially small.

National adversaries are today in many instances hybrid threats, that manifest criminal and military traits, these often require deployment of hybrid-capability autonomous weapons imbued with the capability to taken on both Military and/or Security objectives. The Westgate Terrorist Attack of 21st September 2013 in the Westlands suburb of Nairobi, Kenya is a very clear manifestation of the hybrid combat scenario that required military response and police investigations against a fighting cell of the Somalia based globally networked Al Shabaab terrorist group.

A theoretical solution by way of an Architectural Framework is rendered as a viable solution in this monograph. It seeks to eliminate the practice of procurement without empirically supported deliberation. Inappropriate procurement causes financial losses and inefficiency in tactical operations when a weapon system is fielded, with the overall result being loss or diminishing of sovereignty. Notably the most expensive and painful human endeavours are unplanned military/security operations. Weapons systems must be built with a wide range of operating environments and contingencies in mind, this can only be achieved if ample research precedes the design and development stages.

**Acknowledgments**

I am grateful to the reader/reviewer David K. Ngondi

## Dedication

To my beloved country the Republic of Kenya for all the sacrifices we have made to keep you safe, fully aware that the costly price of freedom is made bearable by way of timely preparation : *Security is a Lifestyle it is not a series of Temporary Measures*.

"**Isaiah 54 : 16**
Behold, I have created the blacksmith who fans the fire of coals, and brings forth a weapon for its work; and I have created the destroyer to destroy"
so proclaims the Almighty YAHUEH, HalleluYAH Amen and Amen

combined Hebrew to English Translation of
The Hebrew Tanakh
according to the
Masoretic Texts and Jewish Publication Society Edition of 1917 [27]
and
Personal Translation

# Chapter One

## Introduction

An autonomous weapon may be mobile or stationary, it is distinguished from manned weapon system types by way of pre-loaded heuristics to undertake its C4ISTAR(Command, Communications, Control, Computers, Intelligence, Surveillance, Target Acquisition and Reconnaissance) and target interdiction functions.

Preloaded heuristics in an autonomous weapon system is designed to process data-streams from its sensors and to output a data-stream that induces a physical response in its actuators, that interact with the operational environment ( Figure A – Appendix A: The Exploratory Study). Automated control systems architecture must specify the sensitivity of its sensors, the nature and data-stream from its sensors and the desired possible actuated response ranges.

Inevitably there are many forms of electronic/electrical "noise", that are found on a system and prevent its perfect functioning. These errors are minimized or eliminated by way of processing of control system feedback, thereby ensuring that original control objectives are obtained. Alternatively filter systems remove the disturbances that are electrical/electronic "noise" from a circuit.

Of greatest concern to a weapon systems engineer are exceptional interference to control objectives as a result of :
1. hardware malfunctions
2. computational errors
3. inadequacies in pre-loaded heuristics
4. jamming by a third party
5. hacking or any for of subversive control

In this monograph the automated control and information systems described are of tactical types for military/security operations. They are robots of many different forms and functions in warfighting, that are challenged by way of dynamic and hostile environments, in addition to exceptional unforeseen occurrences

Norms, exceptions, instability and environmental parameters are obtained by sensors and channeled as data-streams to an automated controlling system that bears the relevant pre-loaded heuristics.

The desired characteristics of feedback or incoming data-streams from sensors:
1. minimum sensitivity that is representative of signal signatures of the operational environment
2. data channels should have sufficient bandwidth/capacity to allow for realization of high fidelity data-streams.
3. Noise must be filtered out of the data-streams wherever and whenever possible

4. Sensors should be specified and developed with the intention of error reduction

Actuators of mechanical types should have degrees of mechanical freedom that are sufficient for reaction to signal data-streams. Some factors considered are:

1. Can the actuators ensure that an autonomous vehicle/robot/device moves to a desired position and in the desired direction?
2. Does the automated system have projectiles, missiles, devices that could be launched in various missions? How far would such projectiles or vehicles travel from the mother-ship? With what accuracy would they execute mission parameters? Are these dependent vehicles also autonomous or are they dependent on the mother-ship for guidance and/or control?
3. How powerful are the actuators in implementation of control decisions? Are they robust or unstable?
4. What are the energy consumption levels of actuators and a whole autonomous device? Significantly low energy consumption rate translates to lower battery or fuel consumption hence a lighter design.

**1.1 Background**

An important consideration in the design, development and fielding of autonomous weapons, is their anticipated missions. These missions are based upon threat modeling and threat characterization the scope of which could include ground, air and/or naval operational environments. For example on the ground environment the anticipated threat could be those such as armed humans or main battle tanks, while in naval operational environment a submersible vehicle/robot may have a specific purpose, e.g. disarming of naval mines. Some autonomous systems could be wholly expendable when they engage a target, while others may be designed in such a way that they track and engage several targets simultaneously before returning to the base.

The focus of the resultant architectural framework is to conceptualize the ways and means for maximizing the desired performance of an autonomous weapon within a wider C4ISTAR set-up. It shall enable the nodes that are autonomous agents to tap into resources of the wider network, while at the same time contributing to the overall situational awareness and actualization of the desired goals in tactical security/military operations.

**1.2 Contemporary Perspectives on Lethal Autonomy**

Autonomous lethal weapons may be categorized into three groups:

Group (1) : Those that are dependent upon the absence of friendly forces or neutral elements in an area of operations if an attack is to succeed.

Group (2) : Those that are dependent upon identification of friend or foe systems borne by elements in a battle-space or area of operations

Group (3) : Those that have humanitarian algorithms and machine learning hence the ability to respond to unforeseen contingencies in the battle-space thereby preserving/protecting friendly or neutral elements in the battle-space.

With these designs there are additional specifications that are borne by a lethal autonomous weapon to assist its owners in obtaining its tactical objectives, e.g.:

1. Autonomous nodes that act in conjunction with other autonomous weapons
2. Effective and robust navigation systems that are equipped with varying levels of redundancies
3. Centralized command centers monitoring various fielded autonomous weapons
4. Autonomous systems that provide commanders with broad-spectrum situational awareness
5. Autonomous systems that can act in collaboration with other autonomous weapons and/or fielded forces
6. Autonomous weapons that have concise awareness of their tactical objectives and the evolving operations area
7. Autonomous weapons that can survive attempted seizure or destruction by adversaries

Given the network integrated nature of many autonomous systems, there is always the risk that cyberwarfare shall be the ways and means of subverting [3] many such weapons, especially if they are dependent upon an open or easily accessible communication medium. This perception by system designers is the basis of what may be known as an ISOLATION POLICY.

**THE TENETS OF ISOLATION POLICIES**

Threats of cybercrime and cyberwarfare [4] must be considered when designing the data-links of autonomous weapon systems to command centres, these are:

**Rule 1 :** Autonomous warfighting robots shall not have provisioning for interactive remote use, resetting or control from mainstream networks, communication media or communication protocols

**Rule 2 :** Feedback data-streams from autonomous warfighting robots shall be only one way type from the robot/weapon to the Command Centers.

**Rule 3 :** Encryption of all feedback to command centres from autonomous robots shall be by way of predetermined one-time key sets for digital encryption.

**Rule 4 :** Autonomous warfighting systems must reduce and/or eliminate their operational reliability via data-links that are based upon degradable communication infrastructure

**Rule 5 :** Isolation of systems blueprints from the wider Internet, that is the work domain of hackers and others involved in theft of data or trade of the same.

Isolation policies ensure that military/security forces that use autonomous warfighting robots, isolate them in a technically sufficient way to reduce or eliminate their dependence of publicly accessible communication networks. The embrace of "isolation" in autonomous warfighting robot networks, may contradict developmental objectives of certain open architectural frameworks in the long run. However they offer some level of assurance that the reliability of critical autonomous robot warfighting networks shall not be compromised by way of a cyberattack from an adversary [5].

Remote controlled weapons were used by Germans in World War I when remote controlled electrical motor boats were deployed for littoral area C4ISTAR duties [6]. These were not autonomous weapons, notably the challenge of use autonomous weapon system comes about primarily when the systems are deployed without 'a human in the loop' – this introduces the perception of moral disengagement of the combatants (who can be the user, the buyer, programmers, analysts, commanders, etc.). To this extent some schools of thought propose that such systems only be utilized for defensive purposes.

There are many International Humanitarian Law rules [7] such as the Geneva Conventions, the Hague Conventions, Rome Statutes, Responsibility to Protect, etc. many of these apply readily in the domain of automated warfighting. The underlying humanitarian concepts therein are:
1. Target identification and discrimination before an attack is carried out
2. Use of military force that is proportional to the threat that is posed by an adversary
3. Ability to abort an attack when target status changes, e.g. a surrender
4. Avoidance of mistreating the wounded

Given the detailed demands of International Humanitarian Law, an autonomous warfighting agent must not only perceive its environment but also detect, designate and track potential targets therein. It must have a way of establishing if it is necessary to engage the target with its weapons. Perception of a target does not only involve physical identification, it also includes determination as to whether the potential target in question is a friend, foe or neutral.

Missions undertaken with use of autonomous warfighting agents may be viable if it is equipped with a wide range of sensors, such as: millimetre wave radars, thermal imaging, acoustic signature detection, seismic sensors, olfactory sensors, etc. The process of targeting would take the following stages:
Stage 1 – Search
Stage 2 – Target Acquisition
Stage 3 – Target Designation
Stage 4 – Target Tracking
Stage 5 – Weapons 'Locking on Target'
Stage 6 – Weapon Launching and Target Interdiction
Stage 7 – Battle Damage Assessment

There is no limit to the extent of applicability of new technological innovations in the domain of Lethal Autonomy. For example with application of principles such as those set out in Lee [8], autonomous unmanned aerial vehicles could communicate with autonomous ground vehicles via projected light. This would allow for the projection of a single message simultaneously to various robots in one geographical area, while at the same time minimize the risk of the message being intercepted.

With such a technique improvements could be rendered by unmanned aerial vehicles communicating with unmanned ground vehicles, using techniques such as quantum optronics to further codify projector messages, to geo-specific projection areas while at the same time minimizing problems such as projection area boundary region data corruption.

The scope and applicability of autonomy in warfare is broad, its applications are multifaceted, and there is an "astronomically" big potential for innovation.

# Chapter Two

# The Architectural Framework

Architectural Frameworks [9] in the domain of military/security, ensures that weapon systems obtained(by way of design and development or purchase) meet and/or supersede the specified mission requirements. Weapon systems are too important to be procured by way of hunches, intuition and impulse, and the consequences of their malfunction and under performance has a detrimental effect on their user's national sovereignty. It is prudent to obtain empirically justifiable reasons to back any weapon systems procurement decision.

Mission requirements for a weapon system include concepts such as: logistical support, reliability, durability, survivability, interoperability, etc. Architectural frameworks allow for comprehensive evaluation of weapon system performance and make all components and aspects borne on such systems easy to subject to an audit/evaluation in the event of an accident, malfunction, or under performance.

Frameworks enable an establishment to compare and qualify systems offered by different suppliers in response to a particular mission requirement. Over the systems service lifespan, Architectural Frameworks could be re-stated to facilitate for the implementation of upgrades, retrofits and maintenance, ensuring that the system in question remains relevant/compatible with ever changing mission requirements.

**2.1 Threat Characterization**

Weapons are tools specifically for neutralizing threats of aggression against our beings or interests. The likelihood of vulnerabilities being actualized by threats is known as the risk. Many states today find themselves confronted by hybrid threats [10] that deploy regular(conventional) warfare methods in addition to irregular(terrorism, crime, asymmetric, illicit trade, etc.) methods. This implies that to confront an adversary of the hybrid nature, a state must deploy both military means and policing/law enforcement methods – this increases the complexity of autonomous lethal weapon development due that the fact that there is not only the requirement of compliance with military law and battle-spaces but an additional requirement for compliance with criminal law and an appreciation of day to day social norms.

A prominent example of a hybrid threat in the region of the Horn of Africa is the Al Shabaab Movement based in Somalia. Other international examples of hybrid threats include the Taliban in Afghanistan and Pakistan, the Revolutionary Armed Forces of Colombia, etc.

Some of the issues that create the felt need for lethal autonomous weapons are:
1. The calculation that experienced warfighters are not worth exposing to the risk of death in the hands of less experienced persons in the course of military/security operations

2. Awkward position of responding forces in a battle-space in cases where an adversary initiates hostile actions
3. Greater propensity of commanders to accept risk exposure of autonomous lethal weapons that are searching and securing a building hence the likelihood to end a siege/hostage situation much faster than a human-fighter
4. An autonomous weapon system can acquire, retain and exploit total knowledge of a battle-space layout

Even with the above listed tactical advantages there are the challenges such as:
1. Non-combatant identification
2. The ability of autonomous robots to rescue hostages, e.g. to carry away children from a hostile environment, win the confidence of victims, handle infants, give first aid to injured persons, and remove injured persons from hostile environments
3. Autonomous conduct of hybrid human and autonomous robot tactical operations
4. Energy issues : battery/generator types, fuel and/or battery energy densities, etc.
5. Mean Time Between Failures

Threats are not only perceived by an autonomous system, they are borne and created by the same. Intrinsic threats are the most difficult to handle as they are not detectable by the autonomous platform in most cases. Extrinsic threats may be characterized in the following fashion:
1. Frequency of their effective projection and execution by an adversary
2. Consequences of their execution e.g. fatality, lethality, incapacitation, etc.
3. Distance from which they can be projected and executed to achieve an adversary's operational goals
4. Ready availability/existence of the threatening entity
5. Vulnerabilities of the threatened
6. Ability of a threatening entity to impair critical functioning of targeted entities
7. Survivability of a threatening entity, its ability to use counter-measures against attack or to evade attacks
8. Level of sophisticated reasoning and planning borne by a threatening entity

For entities engaging in irregular warfare such as terrorists [11] there are many more factors to be consider by the programmers of lethal autonomous weapons that operate against them. Some of the issue to be considered are:
1. Ability of an organization to carry out its activities in multifaceted environment(eg. Air, ground and sea)
2. Mental and doctrinal inclination of combatants
3. Ability of a threatening entity to implement an unanticipated and extraordinary conversion of an apparently 'innocent' item to a lethal weapon, e.g. the use of civil airliner as missiles in the 9/11 attacks in New York
4. Organization structure and discipline or the lack of the same

5. Profile of targets and victims who are attacked or terrorized

Even with all their characteristics threats do not just occur at random, they are developed and actualized by persons who have specific knowledge, intention and resources. These are persons who are of interest to developers of lethal autonomous weapons, as they may constitute a hybrid threat aspect that demands for computer algorithms and machine learning capabilities for irregular warfare.

## 2.2 Systems Architecture Framework

Information Systems including those that are at the centre of autonomous lethal weapons must comply to Local and International Legal requirements. There may be an exception to this requirement as pertains to Information Systems controlling lethal autonomous weapons, that are owned or used by Intelligence Agencies which have quasi-criminal organizations and are only loyal to their paymasters. There are many detailed expert opinions on issues pertaining to International Humanitarian Law, Information Systems and lethal autonomous weapon. Schmitt [12] gives a perspectives that has now become widely accepted within the ranks of the United States of America Armed Services. It lays out details of characteristics, requirements and consequences of lethal autonomous weapons and their controlling systems.

The pursuit of "smart" systems in warfare is not unlimited but constrained, in the recommendations of McCullough *et al* [13] on Strategic and Leadership Recommendations there is the strong opinion that "strong artificial intelligence" should not be deployed on unmanned systems to the extent that resultant weapon systems can learn of their own tactical decisions and make completely independent final decisions without any pre-loaded constraining heuristics. This view is rendered to the effect that "strong artificial intelligence" would result in completely autonomous and unpredictable weapon systems and situation that cannot be allowed to occur by any responsible Commander.

Lin, P, *et al* [14] was an early attempt to specify without codifying the algorithms, heuristics and/or modules that should be borne by a well-engineered information system that is at the core of any wholesome lethal autonomous weapons system. Upon review the key process identified in the document may be laid out in a flow-type diagram, as in Figure 1 on the next page.

Figure 1 : Key Processes in a Generic Autonomous Lethal Weapon System (Intuitively Labeled)

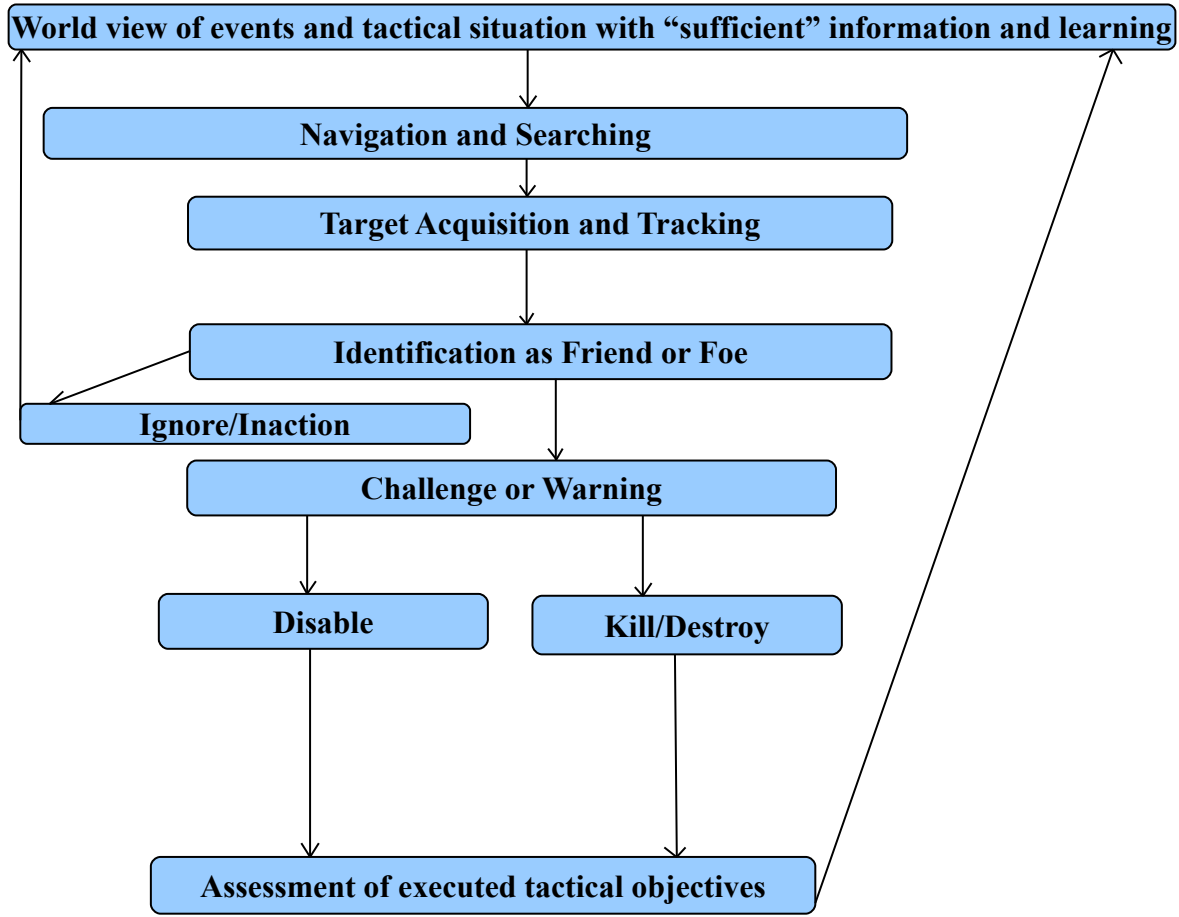

The higher the frequency that a lethal autonomous agent can go through the processes in Figure 1, starting with its world view – the faster its execution of its "Observe, Orient, Detect and Actions" Loop. A faster OODA loop implies that a system is more effective when fielded against to similar system but with a lower OODA loop frequency rate while performing the processes laid out on the flow chart.

Another important parameter would be the size of the "world view" of the information system, how much relevant information does it have? Lack of sufficient "world view" dimensions would render the usage of the system to very limited situations, while too expansive a "world view" would imply that a system has been implemented using "strong artificial intelligence" and is able to independently set out its tactical objectives obtain them and determine its own mission parameters. Complete autonomy of the "strong artificial intelligence" kind is undesired in the realm of International Humanitarian Law for reasons that systems of such nature would be unpredictable and potentially lethal or damaging to the owning military force.

Systems development reviews are important in the course of implementation of a Systems Architecture, they guarantee that an autonomous agent shall be tested and realistically calibrated in an operational environment that is similar if not the same as that of its eventual deployment. Compliance with International Humanitarian Law in systems heuristics is also improved and secured by pre-deployment systems reviews.

Additional codification efforts should be undertaken in the domain of the autonomous robots' perception. Their "world view" must not only be extensive but also fine grained and intensive. The robots must:

1. Recognize human beings
2. Recognize human interests
3. Recognize local and "global" implication of its actions
4. Subvert or override adversary systems
5. Recognize an adversary's threat potentials
6. Analyze available options and possibilities
7. Determine the available ways and means in relation to dynamic mission requirements

There is a possibility that a lethal autonomous system that has "weak artificial intelligence" could develop "strong artificial intelligence" on its own accord if it is endowed with sufficient Genetic Algorithm Mechanisms and Machine Learning Capabilities. In this regard a system that is fairly predictable and well tuned to adhere to the tenets of International Humanitarian Law may become self "willed" and unpredictable in the long run. To prevent the possibility of self-evolving rogue autonomous lethal weapons, a requirement must specify the use of fixed/robust algorithm rule sets in addition to machine learning technologies on a platform. All said and done there are some common military concepts that are not easy to codify into an information system e.g. acknowledgment of commanding authority, proportionality in combat – avoiding overkill, and justification of missions.

Schmitt and Thurnher [12] demonstrates that the legitimacy of military technologies from the perspective of International Humanitarian Law is based upon the prevailing technologies and their availability at the time being considered. Technology that may be discriminate and proportionate today may very well be considered to be an indiscriminate overkill in the future. To ensure compliance with International Humanitarian Law autonomous lethal weapon system require continuous legal reviews before manufacturing, during manufacturing, during use and during maintenance. Without effective legal reviews a simple software upgrade in control heuristics of a lethal autonomous weapon could very easily result in a completely or partially illegal autonomous control system.

Docherty [15] offers a similar approach to design and implementation of autonomous lethal weapons. It suggests the Arkin's Ethical Governor which would be implemented on binary checks on the Laws of War and Rules of Engagement. A weapon system would thereafter be controlled by way of statistical and sensor-based algorithms, that would determine the likelihood of target interdiction within the confines of International Humanitarian Law.

On the issue of Command Centres for lethal autonomous robots, the following capabilities would be required:

1. empathy with those who are fielded
2. knowing when, how and why tactical objectives would be obtained
3. rationalization of ways and means for obtaining tactical objectives
4. verification and assessment of execution of operations for obtaining tactical objectives
5. collaboration of allied forces
6. justification of operations and giving superiors and the general public briefings
7. procurement, deployment and maintenance of new systems

Though "strong artificial intelligence" is yet to be accepted as a means of on-board control of lethal autonomous weapons, it would serve as welcome addition for use in their related Command Centres. "Strong artificial intelligence" would boldly put to a military Commander difficult questions, that may results in inconveniences that would not be obtained if they were occasioned by a sub-ordinate officer. An information system would place appropriate questions to military/security commanders on : the "what ifs", probable loopholes and pit falls, as well as challenging resolutions that may be fallacies stemming from group thinking.

The Command Center would be easily overlooked by the supply side of a lethal autonomous weapons – specifically because of the presumed independence of such a weapon. But the wider objective of deploying lethal autonomous weapons is for reasons of benefiting specific military/security organizations of the interests of the wider society that they represent. For those reasons it would not be tenable for a military/security organization to haphazardly field all manner of lethal autonomous weapons that cannot be monitored via their on-board reporting mechanisms or tracked or co-ordinated accordingly.

A Command Center inevitably brings about the issue of the data-links that enable it to control and co-ordinate various field entities. The specification frameworks for such data-links is laid out on the Technical Architecture section of this monograph.

Data-link requirements are laid out on the basis of available technologies and System Architecture, the key considerations are that the data-links must facilitate transfer of generated information from the lethal autonomous robot to the Command Centre. Information required by commanders and generated by a weapon would include navigation, targeting, rate of consumption of on-board resources, malfunctions, sensor data-streams, etc. Given the probable likelihood of Information Overload at a Command Centre, "strong artificial intelligence" may assist commanders in fully exploiting what may be difficult to discern and high tempo information – these are likely to be "wasted" if left for handling by way of unaided human capabilities.

## 2.3 Technical Architecture Framework

A wide range of standards and conventions ought to be actualized if truly lethal autonomous weapons are to be realized. There would be the initial hardware platform and software environment issues, these would be at the core of the system. An autonomous lethal system would be required to address a certain level of Problem Complexity – depending on the technology available for on-board computing, this would have a direct impact on the weight, dimensions, functionality, and size of the weapon in question. Some basic technical modules on-board a lethal autonomous weapon would be:

1. guidance and navigation module, eg. Auto-pilot,
2. power and resources management module
3. application/user interface module, for reprogramming and maintenance at the home based of via remote control.
4. targeting/tactical module
5. communications/data-link module
6. self-diagnostics and self-healing modules
7. electronic counter-measures module
8. electronic counter counter-measures module

Technically speaking a more sophisticated autonomous weapon would have more technical modules of varying complexity. There are however great limitations on the ability of many anticipated lethal autonomous weapons to carry on-board the requisite computational power that would allow for the resolution of exponential time computational difficulty type problems within desired time ranges. These functions would inevitably be resolved by computation power that would be resident at a Command Centre, this technical challenge adds to the burden of developing a secure and tamper proof data-link to facilitate full and proper functioning of a lethal autonomous weapon that utilizes off-board computational power in some cases.

There shall never be sufficient computational power on-board a lethal autonomous weapon and the temptation to source for more computation power externally shall create both technical and information security challenges for systems architects. Relative to advances in the scientific world even if all the computational power in the world today could be compressed onto an on-board platform of some military robot, by the time that such an advance became possible, there would be even more external computational power available for lethal autonomous weapons.

**Figure 2 : Key Technical Standards required in a Generic Autonomous Lethal Weapon System**

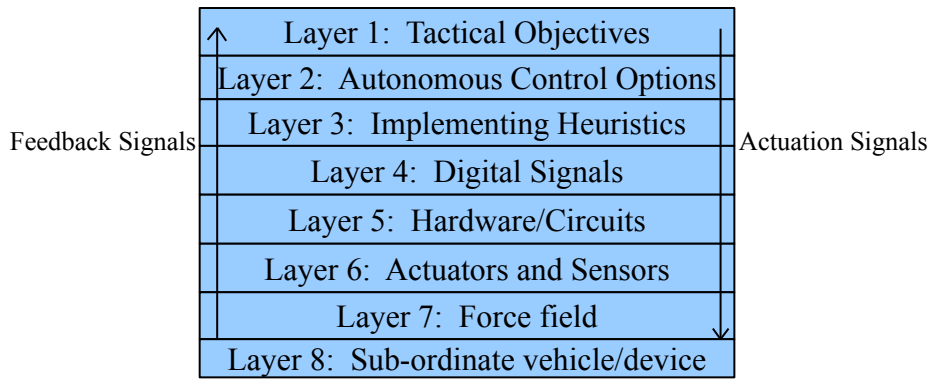

Mother-ship Layers : Layers 1 to 6 are found within the primary device/vehicle that is why they are classified as the "Mother-ship" Layers.  Layer 7 is the "Force field" layer it comprises of radio-frequencies, audio sonar signals, lasers, gases, liquids and/or substances, that are projected into the Operational Environment by the Mother-ship and are used to make measurements/detections of various kinds.  The final layer comprises of sub-ordinate vehicles or devices that are borne by the Mother-ship and deployed in the operational environment outside the Mother-ship and are manipulated, control or communicated with via the active Force field layer.  Sub-ordinate vehicles/devices may be a swarm that works in synchronized fashion via artificial intelligence or a collection of vehicles/devices that work in tandem as a signal relay.

An example of operations conducted by sub-ordinate vehicles/devices would be an unmanned aerial vehicle dropping small crawling robots on to the ground for target verification and assessment.  Given that some of the target may be inside buildings, the crawling robots may position themselves such that some go into the building and some remain outside the building as signal relays.  Once inside the building a crawling robot would take a photo, DNA sample, etc. and send the results via a signal relay of other crawling robots to the Mother-ship that is an unmanned air vehicle.  There is no limit to this concept as a sub-ordinate vehicle/device could also be a Mother-ship to other sub-ordinate vehicles/devices.  The selection of data-links for these multifaceted operations would depend of factors such as energy densities of battery technologies, prevention of jamming and interference, etc.

Functions of the Layers:
Layer 1 : Tactical objectives the desired outcomes issued to an autonomous lethal weapon by its owners/users
Layer 2 : Automated control options are the suggested methodologies for obtaining tactical objectives
Layer 3 : Implementing heuristics select and execute the most appropriate automated control decision
Layer 4 : Digital signals are generated to actualize the implementing heuristics
Layer 5 : Hardware generates and bears the digital signals that are convey from the sensors and to the actuators
Layer 6 : Sensors perceive the environment while actuators manipulate the environment
Layer 7 : Force field are aspects of the operational environment that are manipulated by the autonomous platform

   then monitored for purposes of surveillance or control

Layer 8  : Sub-ordinate vehicle/device is launched from, controlled and/or monitored from the Mother-ship for purposes of enhancing its off-board capabilities within the operational environment

Some level of independent decision making is expected from an autonomous lethal weapon, this implies that an adversary must develop a specific attack type if the autonomous lethal weapon system is to be subverted. These specific attack types are directed at specific layers of the system, hence a requirement that specific layers must have specific in-built counter-measures. The following are some attack types vulnerabilities that an autonomous weapon system designer should anticipate and work to prevent on the fielded system:

Layer 1 Attacks :  Attacks against programmers and commanders, these can be physical or mental
Layer 2 Attacks :  Sabotage embedded into the system inform of hardware, software, firmware at procurement
Layer 3 Attacks :  Malicious code introduced via data-links into the system
Layer 4 Attacks :  Electromagnetic pulse attacks and microwave attacks
Layer 5 Attacks : Kinetic and non-kinetic direct attacks against sensors and actuators
Layer 6 Attacks : Implanted counter-measures(sabotage) – credibility problems within ranks of system users/owners
Layer 7 Attacks : Jamming/interference, decoys, etc.
Layer 8 Attacks : all possibilities from Layer 1 Attacks all the way to Layer 7 Attacks

There are various standards to protect systems from these threats, organizations such as North Atlantic Treaty Organization, International Organization for Standardization, Institute of Electrical and Electronic Engineers, United States Department of Defence standards, etc. these would be applied either at the point of systems design and development or used for verification at the point of procurement decisions. It would be improper to give specific standards in this monograph as it is a framework and not a definitive view for any autonomous weapon system.

With standards come the additional challenges of testing and calibration, verification and validation, once as system is deployed there is the continuous practice of monitoring and evaluation. In practice not only are current threats and issue for concern but new and emerging threats that may render an autonomous system obsolete or inefficient in its domain of operations. Various retrofits of sensors and actuators are undertaken, in addition to upgrades of both hardware and software components. Upgrades may be easily undertaken if an autonomous weapons system was built in modular fashion, such that older components can be unplugged and new components plugged in, followed by rebooting and reconfiguration.

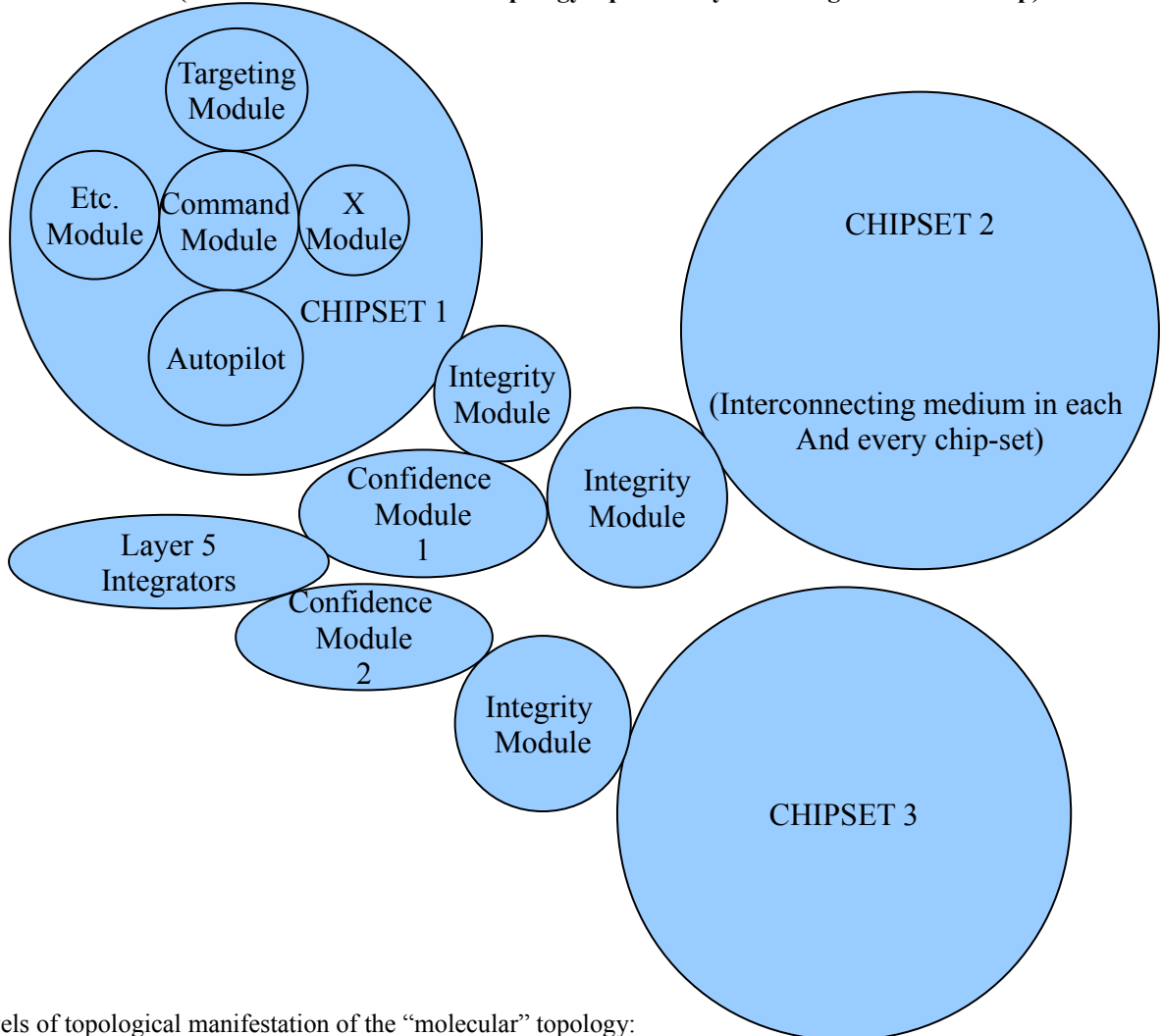

**Figure 3 : Generic Autonomous Lethal Weapon System – Fail Safe / Fault Tolerance Mechanisms
(Autonomous "Molecular" Topology – preferably on a integrated circuit chip)**

Levels of topological manifestation of the "molecular" topology:

1. macroscopic implementation
2. microscopic implementation
3. nanoscopic(eg. quantum device) implementation
4. hybrid(combination) implementation

Macroscopic implementation : In typical macroscopic implementation you can see the actual capacitors, resistors, transistors, and logic/controlling devices

Microscopic implementation : on chip based integrated circuit format

Nanoscopic implementation : too small for normal electronics instead quantum level manipulations occur.

Hybrid implementation : a combination of macroscopic, microscopic and/or nanoscopic circuit types for to meet a specific design requirement or to overcome the challenge of procurement of rare or unavailable components.

The key concept behind the "molecular" topology is that system failure in one domain would not result in overall system failure which may have catastrophic effects. For example if one auto-pilot module reads questionable data, other auto-pilot modules designed in different ways may read different data altogether. The confidence modules would then allow through various reading that would be worked on by the "integrator" which would in effect only accept operational parameters that are of very high confidence levels, strange or outlier parameters shall be discarded.

Another approach would be for example if an auto-pilot module in one chipset fails completely, data and functionality could still be obtained from other auto-pilot modules in the same or other chipsets. No function or capability of the autonomous weapon would depend upon a single on-board component leading to a robust and highly dependable system. To further improve the survivability of an autonomous lethal weapon platform, the "molecular" topology chips would be located on different parts of the platform, each with its own range of modules and a network of integrators to read their confidence levels all through the platform – thereby only executing the most appropriate parameters.

In the "molecular" topology a chipset performs some function(s), the integrity modules then filter out absurd or erroneous data from further consideration and application within the system. After filtration of outlier data/parameters, the confidence module compares data-streams from a wide range of chipsets and gives a determination as to the level of confidence based upon its scoring frequency. For example if several confidence modules working in synch make the determination by 89.9% that a vehicle is on the equator, then it is overwhelmingly probable that it is in the defined navigation area. But levels of confidence may depend on the sensitivity of an operation, e.g. in some operations a 50% level of confidence may be acceptable while in other operations a 95% level of confidence may not be sufficient. In future designs a chipset shall simulate brain neocortical column [24].

Data-streams that are marked via encapsulation as being sufficient are thereafter utilized for execution of actuator in navigation, control decision and tactical operations. The implementation of "molecular" topology is not fixed and could take many other forms and components as its statement in this monograph is purely at conceptualization and basic functions level. More complexities would inevitably crop up in the pursuit of "molecular" topologies for lethal autonomous weapons design, the most difficult would be the development of software for running the systems which in many cases would be proprietary due to the secret nature of their development.

## 2.4 Operational Architecture Framework

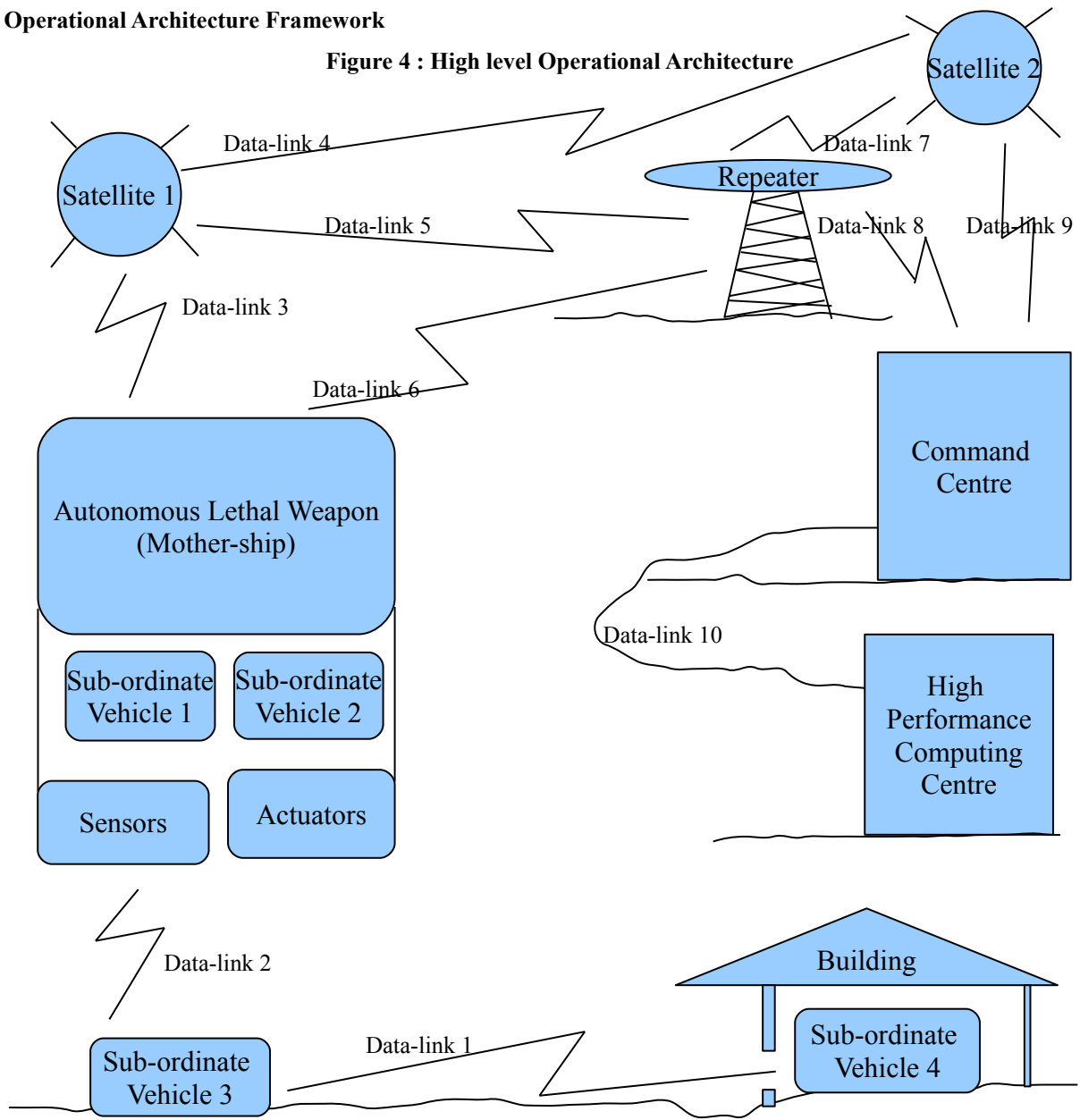

Figure 4 : High level Operational Architecture

In the Appendices, The Exploratory Study page 6 of 15 there is Figure A, that is a schematic of the middle-level Operational Architecture view of an autonomous lethal weapon system, Figure 4 gives a higher-level Operational Architecture view of an autonomous lethal weapon system (Mother-ship) that has deployed sub-ordinate vehicle 4 to search inside a building and to relay its findings via the electromagnetic spectrum to a sub-ordinate vehicle 3 which then repeats the signal that is strong enough to be received by the mother-ship. The mother-ship relays its signals via terrestrial radio-frequencies repeaters or satellite data-links to the command center for monitoring and evaluation.

The mother-ship has on-board two more sub-ordinate vehicles that it can deploy for similar purposes, these extend its capabilities of target search and interdiction as without the sub-ordinate vehicles it would not be in a position to discover or verify the status of persons or objects in a building. The set-up would further enhance enforcement of International Humanitarian Law by way of reducing civilian/non-combatant casualties, allow for better assessment of proportional use of military force and if well designed it could assist in the conduct of physical battle damage assessments, that have previously been only possible by way of aerial and satellite photography. The design assumption is that a sub-ordinate vehicle is equipped with the appropriate actuators and sensors – there is also the possibility that a sub-ordinate vehicle could be the mother-ship of another type of sub-ordinate vehicle.

Even with technologies that are currently in the market the various data-links can be specified for purposes of demonstrating the practical nature of this architectural view and not for reasons of placing a definitive specification. There may be the restriction of attempting to process as much data as is technically possible on-board an autonomous platform and to utilize only one-way half-duplex communication to reduce and/or eliminate the risk of hacking and seizure of an autonomous platform by third parties. A sample specification of the data-links at a high level may be as follows:

Data-link 1 : Links sub-ordinate vehicle 4 to sub-ordinate vehicle 3 → eg. Wireless fidelity

Data-link 2 : Links sub-ordinate vehicle 3 to mother-ship → eg. Very high frequency radio

Data-link 3 : Up-link from mother-ship to satellite 1 → eg. Optronic link of lasers with quantum encryption

Data-link 4 : Link from satellite 1 to satellite 2 → eg. Optronic link of lasers with quantum encryption

Data-link 5 : Down-link from satellite 1 to terrestrial repeater → eg. Optronic link of lasers with quantum encryption

Data-link 6 : Radio link from mother-ship to terrestrial repeater → eg. Very high frequency

Data-link 7 : Up-link from terrestrial repeater to satellite 2 → eg. Optronic link of lasers with quantum encryption

Data-link 8 : Radio-link from terrestrial repeater to command centre → eg. Microwave

Data-link 9 : Down-link from satellite 2 to command centre → eg. Optronic link of lasers with quantum encryption

Data-link 10 : Command centre to high performance computing centre → Fibre-optic cablelink +10 Gigabits/sec.

A wide range of systems engineering decisions can be made in actualization of this architecture, e.g. transmission of video and graphics in monochrome to reduce bandwidth and/or utilizing of schematics in place of videos. Mission packages could be made such that particular tactical operations only have the required type of data-links and processing capacity on the mother-ship to increase its endurance without refueling or replenishment of any kind. Energy replenishment is a particular problem in the domain of unmanned ground vehicles, unlike the human warfighter who can live off the land with minimal resources, the unmanned ground vehicle has got the challenge of fuel supplies and battery life. Human muscles are highly efficient structures, apart from being a vital part of the actuators in the human body they also store energy for use – robots on the other hand need batteries or fuel tanks.

With an unlimited supply of power a robot fielded in combat may transform itself into a different entity or to use its power pack for camouflage illusions. There are many things that a human fighter can do that many of the fighting robots in the market today cannot. For example a human-fighter may create relationships with adversaries or non-combatants, yet a robot may not achieve the same. The human-fighter is able in most instances to carry away wounded persons and to win their confidence, a robot may also fail in this respect – a robot may have the power to carry the wounded but may lack the sensitivity to avoid inflicting more injuries. Other similarly challenging operations for a robot may be those such as picking locks, using objects found in its environment as weapons, breaching walls and doors, etc.

Weizman [16] details the views of two Israel Defense Forces commanders Brig. Gen. Aviv Kokhavi and Brig. Gen. (Rtd.) Shimon Naveh, pertaining to limiting exposure of ones forces to enemy fire during urban warfare. It advocates for pursuit of tactical objectives in a city or town without going through readily available routes. The result of this advice is breaching of new routes through buildings with use of sledge hammers, explosives, cutters, etc. in a process that is dubbed as Lethal Theory. These innovative approaches to urban warfare are a challenge to replicate in the autonomous realm with readily existing technologies as of the year 2013. Ability of robots to swing sledge hammers to break walls, breach roofs and floors with impunity, would require energy densities in robot power sources that allow for the equivalent of heavy duty human work or surpassing the same.

Weizman [16] brings in the concept of full and unhindered autonomous movement in the urban realm, to the extent that a warfighter is able to breach any obstacle/impediment to movement. With fully autonomous movement capability a warfighter would easily outmanoeuvre adversary forces in the urban realm as they would not expect to be attacked from a new direction of approach that did not exist prior to the fighting. An autonomous lethal robot designed to implement "lethal theory" should have the ability to interpret architecture and civil structures, without restrictions of avoidance of walls, floors, roofs, and other non-transit areas. At the same time progress of the robot would be reported to high command via some implementation of high-level operational architecture. In effect an efficient autonomous lethal robot would not comply with an adversary's expectation for pre-specified transit route.

Each and every data-link relaying commands and information to and from the robot/weapon requires specification for a particular range of tactical operations. Specification of interest would include but not be limited to : Transmission power; type of computing processes, actuators and sensors that are communicating with off-board systems – as these would describe the information being exchanged; security and counter counter-measures implemented in the data-link; amount of information exchange required internally and externally for efficient operation of the robot/weapon; tempo of work that is generating the information; rate of consumption of the information by external systems; ability of robot/weapon to establish data-links with other systems to facilitate interoperability; etc. Note: the greater the information throughput the higher the energy consumption

### 2.5 Potential Applications

In the course of research work of this monograph the Westgate [17] Terror Attack occurred in Nairobi, Kenya on 21[st] September 2013 and lasted for four days. A group of about 8 to 15 armed gunmen engaged the Kenya Defence Forces and the Kenya Police in indoor fighting at a Shopping Mall located in Peponi Road, Westlands Area of Nairobi. Not only did the military and security forces engage in indoor fighting but they also had to secure the wider area to prevent escape or reinforcement of the gunmen. There were a wide range of challenges that were encountered in the course of the fighting that may be of interest to a developer of a lethal autonomous system, eg.:

1. is there an autonomous robot that can be deployed to carry out tricky operations such as first aid, rescue, negotiations, etc.?
2. could the deployment of autonomous robots assist military/security forces to reduce the risk of loosing highly experienced and expensive to train warfighters in the hands of less experienced persons?
3. what are the threats and opportunities in conducting tactical operations that involved autonomous robots and human warfighters?
4. Autonomous robot sentries may guard valuables such as jewels, cash, food, etc. from exploitation by the terrorists while other aspects of the tactical operation are conducted. They would be less inclined to misappropriate would they use or remove, and would keep a digital record of the same
5. Networked autonomous weapons could identify adversaries via long range facial recognition or by firing projectiles that are sub-ordinate devices that can embed themselves on an attacker and report his/her Deoxyribonucleic Acid profile. Allowing for positive identification of adversaries before they are killed or captured.
6. The ability of an autonomous robot to search the building would be much faster due to its greater propensity to expose itself to risk of destruction or damage, than human willingness to expose oneself to harm/injury or death when searching a building or other battle-space type
7. Ability of an autonomous robot with in-built electronic warfare suite to intercept and exploit terrorists communications via phone or radio. During the siege, security/military forces did not have full exploitation on the electromagnetic spectrum, ie. They did not have an 'Electronic Order of Battle' mapping out the electromagnetic signals in the vicinity. In addition to 'identification friend of foe systems' an electronic warfare suite in an autonomous robot may end a siege faster by tracking down adversaries as they use radio-frequencies and prevent incidents of friendly fire by detecting presence of friendly forces.
8. An autonomous robot fighting within a city or a building could be plugged into the close circuit television system to give it an augmented perception of the battle-space, in turn the robot may also communicate with the hostages and terrorists via data-links with the building intercom system or the phones or radios that they are using during the incident

There are very many limits to use of the 'Westgate Terror Attack' as a benchmark for feasibility of potential deployment of lethal autonomous weapons. Notably the terrorists were restricted to ground fighting which did not have a naval aspect – but there was an air component of the Kenyan side comprising of helicopters and planes. The fighting environment was primarily urban and indoors. The face-off between the Kenya Defence Forces and Al Shabaab inside "Westgate" Mall was asymmetric in nature – the Kenyan Military was endowed with a home advantage and broad array of technical capabilities.

Ground, naval and air warfare between well equipped adversaries with proper military training, is a ultra-high tempo affair, with gains or losses in the billions of dollars range are quantified by the millisecond and thousands of tactical operation instances are executed in similar time.

More technologies [18] are now available in the market that would facilitate for the development and use of autonomous lethal weapons, an example is the development of Advanced Extremely High Frequency satellites with data-links of 8.1megabits per second and Active Electronically Scanned Antennae technology that makes it easier for ground-stations and satellites to establish connectivity. The advantage of these technologies would be to provide beyond-line-of-sight communication between autonomous lethal weapons and command centres. Though high capacities for data-transmission exist in civilian technologies, they cannot be deployed in remote and unsupported environments, hence the emphasis on bandwidth provision via independent military satellites.

Other innovations that are gaining rapid acceptance in the market are "throwables" [19] that are robots that can be deployed into a building/battlespace via an awkward and unplanned throw. If equipped with appropriate artificial intelligence a throwable robot could search for and locate items and persons, map out buildings, avoid/breach obstacles to allow for the odd "look-around-the-corner". With appropriate data-links throwable robots would promptly furnish their users with audio, video or other data as pertains to conditions in its operating environment.

Throwables by their very nature as robots are rugged and retrievable, on the extreme end of the robot spectrum are the "One-way-Ticket" drones/robots [20], once launched or deployed into a battle-space these use various homing techniques to lock on to target signatures and navigate to their locations before attacking. In the resultant attacks the drones/robots as well as the target are destroyed. There are many drones with this capability such as the Israeli Harpy and the German TARES. Given the nature of these drones, they are usually deployed against high value targets such as radars and command installations.

The growth of tactical internet [21] technologies provide an open standard for connecting to, configuring and manipulating the user interface of a fielded autonomous weapon. Tactical internet enables shared control and exploitation of the on-board resources, via remote control. These are important in control and integrating systems such as the autonomous Kongsberg Crows Weapon Station [22] for Forward base protection used in land warfare.

Lethal Autonomy has spread to highly specialized military roles such as submarine warfare [23] there are various types of unmanned underwater vessels such as the anti-submarine continuous trail unmanned vessel, conceptualized by the Defense Advanced Research Projects Agency of the United States of America. It is capable of trailing a modern silent diesel submarine over an eighty day period and has an operational range of about several thousand kilometres. When tracking an adversaries submarine it is not distracted by other ships and promptly evades them.

**2.6  Conclusion**

Lethal autonomous weapons if well conceptualized, designed and engineered offer tactical advantages to military/security forces that possess and operate them.   Problems such as shortage of skilled military/security personnel could be alleviated to some extent by way of deployment of autonomous lethal weapon systems.  International Humanitarian Law can address the issue of lethal autonomous weapons if and when they are considered as persons/combatants – otherwise  when considered as inanimate objects they could undertake operations such as physical transformation to innocent or non-combatant mimicking entities that are lethal.

With rapidly evolving technologies eg. Further miniaturization of microchips and more efficient actuators and sensors, there is the promise that long endurance autonomous robots, shall be critical actors in today's battle-spaces.

## Chapter Three

## The Findings

In this monograph it has been clearly demonstrated that sufficient knowledge and technology exist for the development of lethal autonomous weapons. Extent of technical details is dependent upon many factors such as:

1. Complexity of components and systems available to a contractor
2. Fine-grained nature of current ethical, legal and tactical requirements
3. Extent of knowledge of systems architects and engineers
4. The ability of a targeted adversary to evade a lethal autonomous weapon and/or deploy counter-measures, i.e. vulnerabilities of an adversary
5. Available C4ISTAR infrastructure for monitoring and/or intervening(during unanticipated contingencies).
6. Nature of the operator of the autonomous lethal weapon, e.g. police, military, criminal/terrorist/non-state actor, intelligence agency
7. Urgency of deployment, i.e. time available for realistic research and development effort

In reality lethal autonomous systems today exhibit a limited scope of heuristics and machine learning capabilities, systems that are not fully capable of anticipating and reacting to all manner of contingencies. A more controversial concept in autonomy would be that of Kantian autonomy [14: pp. 33, 64, 65, 84] if exhibited by a robot would result in pursuit of "best interests", freewill and personal advancement, concepts which are currently only known to be embraced by living organism especially of the human type. Some proponents assume that this state of affairs shall be possible within the next ten years or few more years – when the "Human Brain Project" of Henry Markram [24] and his peers manages to simulate a complete and realistic neocortical column of the human brain.

A simulated complete human neocortical column should enable the computer scientist and systems engineer to "implant" or install some measure of independent thinking into robots. Whether this may be of immediate benefit to human owners of robots is highly debatable, as robot may make the independent decision that are its own robotic best interests and not that of the human owners/commanders. With a model brain on-board an autonomous lethal weapon, many forms of thinking and decision processes become possible as a robot develops and evolves its own sophisticated world view. Decisions would enhance its perception to the extent that there is either negative, positive or neutral feedback from its operational environment that feeds the learning processes of the robot "brain".

Even with the best on-board robot "brain" there is need as in human society for a person to consult experts for greater clarity on complex issues. Lethal autonomous weapons shall utilize data-links of send complex queries to command centres and high performance computing establishments with the expectation of yielding efficiency.

A notion that International Humanitarian Law is sufficient in its current format to meet and address all challenges of lethal autonomous robots in warfare would only hold if it assumed that a robot is not a soldier. If one is to assume that a robot is a soldier many questions would arise as to the legitimacy of the following in conflict:
1. Self assembling robots that could reconfigure themselves into seemingly neutral objects
2. How to deal with the issue of military uniform, insignia or the lack of it
3. Whether a robot can be charged with the war crime of lethal treachery if it pretends to surrender then kills its victims as they attempt to process its surrender
4. Dependency of a lethal autonomous robot on civilian data-links and relays may lead to targeting of the same by an adversary as legitimate military objectives thereby causing human casualties
5. Lack of knowledge of algorithms and hardware that constitute lethal autonomous weapons may automatically render usage of the same by a military as a war crime, given that they would have failed in their obligation to institute a legal review of the said system within the context of International Humanitarian Law

Most of the autonomous robots/weapons reviewed in course of developing this architectural framework have very limited operational scope and functionality. A fully functional combat robot capable of handling high computational difficulty tactical situations and equipped with the required actuators may take a detailed technical specification that is beyond the scope of this research but based on the basic architectural framework described in the foregoing. Tasks and capabilities are primarily in the domain of simple surveillance and their usage is in hybrid mode, support of human tactical operatives, these are not the only autonomous combatants in the market today, there are many gun stations, remote sensing and detonating mines, etc. Unlike the current human operatives the robots are yet to fully develop or be developed such that they can procure, learn to use and deploy weapons of different kinds.

Simple tactical moves are near impossible if not impossible to the current lines of lethal autonomous robot warriors. A gunnery station could run out of ammunition yet be incapable of collecting a rifle of a fallen adversary and deploying the same in combat as a weapon – this is because most robots are designed with specific missions in mind with little or no actuators that mimic or supersede the degrees of freedom of human limbs. These shortcoming are a consequence of current scientific and technological knowledge levels of expert human beings :
1. Energy densities of batteries and mini-generators that power robots, their data-links, processors, sensor and actuators, and the technical efficiency of energy usage by lethal autonomous weapon components
2. Inability of lethal autonomous weapons to carry on-board all the computational power for fully autonomous operations, necessitates the dependence of data-links to external high-performance computing facilities
3. Lethal autonomous weapons are mission specific not designed as tactical competitors to human operatives

Pseudo-human lethal autonomous weapons would fill a design requirement gap herein named as the "H-Gap".

**3.1 Figure 5: The Conceptual Framework – Euler Diagram of Lethal Autonomous Weapon Design Variables**

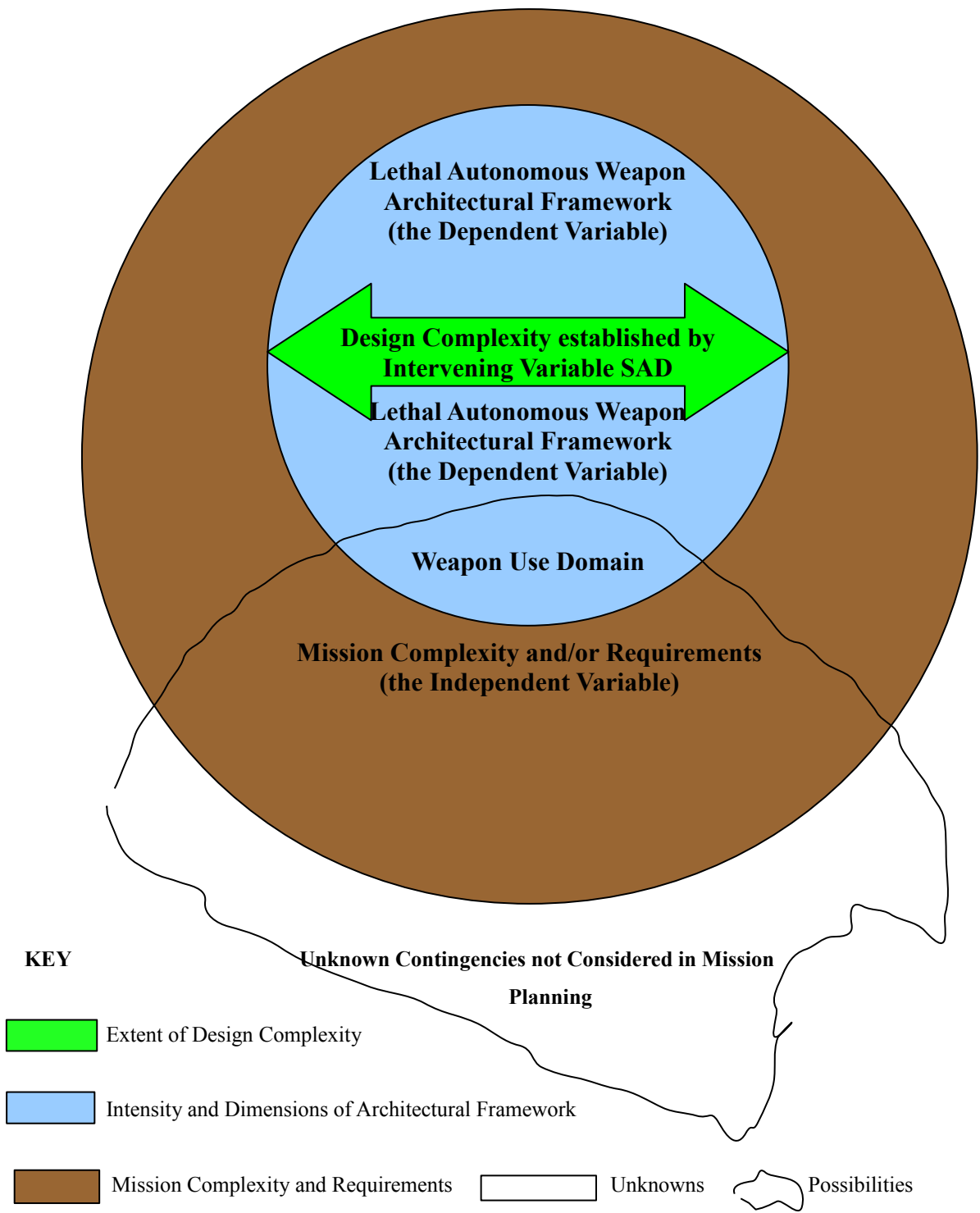

From an intuitive analysis of the conceptual framework [25] above it is clear that there is considerable white background around the circle that is brown, that marks the domain of "everything possible that can go wrong" in a mission undertaken by way of a lethal autonomous weapon. A segment of this white background is further demarcated by way of a non-linear line, this is the space of things that actually go wrong or do not occur as planned during any particular mission. The whole non-linear line segment cutting into the brown circle is indicative of the mission complexity and requirement issues that come up during an mission for which a lethal autonomous weapon is not optimized to react to but it can detect and record. The non-linear line segment in the light blue circle is indicative of challenges that a lethal autonomous weapon encountered and reacted to appropriately due to its design.

The extent of the Architectural Framework is indicated by way of the two tipped light green arrow in the centre of the light blue circle(the Architectural Framework). These are the primary design variables that influence design, engineering/implementation of a lethal autonomous weapon. After a security/military mission, computer programmers and systems analysts attempt to cover an ever wider scope of mission complexity/requirements before the next mission is undertaken – inevitably the extent of the Architectural Framework is expanded. Improvements result in ever greater design complexity, these in turn increase the scope of catastrophic failures that may occur during a mission. An expanding Architectural Framework seeks to deploy more redundancies in a bid to boost fault tolerance which inevitably increases the complexity and types of pre-mission testing and maintenance tasks.

Some schematics capability provided for via a Project Management software application can facilitate for the layout of similar conceptual frameworks in the course of lethal autonomous weapon projects. Such provision would render a visualization that would aid budget allocation for mission challenge/problem solving. Visualization of a project related conceptual framework may aid in the determination of most effective project teams at various stages. In the absence of project management applications the same nature of analytic overview and task visualization could be achieved by way of a Spreadsheet application package. The conceptual frameworks of similar scope and type may also be sourced from information systems for tactical analysis that simulate performance and functions of weapons.

A periodic analysis of conceptual framework scope for different projects but sourced from the same methods and systems would be a pointer to overall project implementation efficiency. Information Systems professionals would be in a position to determine how to optimize the performance of project teams and to minimize wastage and task duplication. In this respect the conceptual framework has got theoretical as well as practical implications/applications. In the above described conceptual framework the variables are described as follows:

1. Independent variables – Mission Complexity/Requirements
2. Dependent variables – Design Complexity based upon the Architectural Framework
3. Intervening variables – Systems Analysis & Design (SAD) work

## 3.2 Discussion

In the "mind and thinking" of a lethal autonomous weapons the Mission Complexity/Requirements are actually a series of Computational problems, that assessed and analyzed with varying degree of Computational Difficulty. The Computer Science domain of Computational Complexity has a role to play in meeting and superseding the challenges of Mission Requirements/Complexity. There is an additional task of discovery of unknowns and contingencies that cannot be resolved by on-board autonomous reasoning capabilities of the weapon and that are therefore sent off-board to command centres and high-performance computing facilities for determination. Secure communication via data-links of these scenarios is part and parcel of autonomous function off-board. The Weapon Use Domain indicates the performance parameters within which the weapon responds to a mission as designed.

Demaine [26] lays out the basic concepts of computing in terse clarity, these are important when considering the "hows and whys" of computation capabilities in future autonomous weapons. Of great interest is his assertion that there are more decision problems than computer algorithms/programs to solve them. An assumption is made that a computer program is a finite string of binary bits, hence its is assumed that the number of computer programs are countably infinite $|\mathbf{N}|$ while decision problems are denoted by $|\mathbf{R}|$ are uncountably infinite. $|\mathbf{R}|>>|\mathbf{N}|$ therefore indicates a special problem for designers of lethal autonomous systems, to the extent that there are ever more decision problems to be solved in an operation than are the number of computation tools/systems to resolve them.

With the $|\mathbf{R}|>>|\mathbf{N}|$ assumption and considering lethal autonomous devices/weapons are in effect systems that undertake to resolve decision problems, it is safe to assume that their designs shall always be inadequate unless and until something beyond normal computing is developed e.g. development of "spontaneous" algorithms to handle the infinite chaos arising from signal noise and relevant signals. These "spontaneous" algorithms would further be mutated by way of genetic algorithms with a view of optimizing their performance. The critical challenge to the concept of a "spontaneous" algorithm would be that of sensor and actuator efficiency, as such partially-efficient or incomplete perception signals from sensors shall result in generation of erroneous and/or hazardous spontaneous algorithms.

"Spontaneous" algorithm solutions are classified as non-deterministic polynomial time problems in Computation Theory. If an autonomous weapon system encounters a NP – hard problem (one that manifests all forms of computational challenges found in the NP problems) or other #P – complete problem(one that is more hard than an NP-complete problem), it would depend on external computational capacity to resolve it. The NP – hard problem would be discovered and packaged before being sent of via a secure data-link to a high-performance computing facility. The size and scope of autonomous lethal weapons contemplated in this paper assumes that they cannot carry such computational capabilities on-board, this shall remain true to the extent that quantum computers in particular are not mobile and/or portable.

Even the seemingly autonomous fielded human soldiers are dependent upon expertise of other persons not physically present in their immediate area of operations, the availability of these requirements for consultation are is accessed via military radio tactical radios – that in many instances offer some form of secure data communications.

**3.3 Conclusion**

New and emerging ideas have the challenge of overcoming the resistance that is the current Establishment. New thinkers often lack the ways and means for propagating their new ideas. With a lot of investment into the human soldier and manned weapon systems, the logic of unmanned autonomous systems has a ready-made market barrier. In general barriers for adoption of lethal autonomous systems technology include:

1. Some commanders may prefer manned systems
2. Lack of trained personnel capable of designing, developing and implementing lethal autonomous systems
3. Lack of industrial base for manufacture i.e. an interested entity may not have access to components
4. Almost intractable engineering problems e.g. energy and endurance of robots in operations, degrees of freedom of movement
5. Autonomous weapons are likely to have very specific objectives and may not wholly "comprehend" the overall tactical situation
6. Determination of on-board and off-board computation capabilities of lethal autonomous systems
7. Data-links and data-communications security against problems such as jamming, interception, alteration.
8. Lack of appropriate technologies e.g. nano-technology, self-assembly, spontaneous algorithm generation.
9. Resistance from human rights activists, skeptics, etc.
10. Fast pace of development of computer technology making it likely that tech-saavy adversaries may be in possession of better performing autonomous weapons
11. Lack of appropriate design and performance bench-marks
12. Lack of testing in real combat situations, appropriate testing situations may take place in parts of the world that are far away from manufacturers. When they occur there is not only the distance-time disconnect but the situations arise in different countries under different cultures and government. This results in the challenge of adoption of the "alien" technology.

To overcome the financial challenge of developing lethal robots some business establishments [29] have resorted to use of investors rather than long-term, government funding contracts. Most of the technology for now works well in urban warfare environment as it is easy to get the components and to draw up appropriate systems integration frameworks. Investor funded lethal robots have development spans of few months and cost a small proportion of those of government funded robots.

Autonomy is still the key consideration of government funded lethal robot projects [30] especially in the United States of America, where it is envisaged that autonomous battlefield ground vehicles shall be operational by 2015. There shall be a time when even the most complex hostage situations, e.g. piracy at sea, shall be resolved by autonomous "frog man" robots.

**IMPORTANT NOTE: Times New Roman, font size 10 and Harvard referencing [1][2]used on this document**

**Appendices**

**Appendix A : The Exploratory Study (on next page) :: undertaken prior to the Monograph**

"Post-Westgate SWAT : C4ISTAR Architectural Framework for Autonomous Network Integrated Multifaceted Warfighting Solutions
Version 1.0" - A Peer-Reviewed Monograph

by

Nyagudi Musandu Nyagudi - Security Analyst

Nairobi, Kenya – EAST AFRICA

**Exploratory Study**

(Exploratory Study preceded the Monograph)

**Reader :**

David K. Ngondi, works at the Directorate of Police Reforms of the Kenya Police Service and has several decades of experience in a broad-spectrum of security and paramilitary operations. He possess a Master of Arts degree in Security Studies from the University of Hull in the UK.

**Keywords:** Informatics, Information Systems, Computing, Signal Processing, Control, Forensics, Military, War, Robotics, Ethics, Automation, Autonomous, Homing, C4ISTAR, C4ISR, Battle-space Digitization, .

**Contents**

**List of Abbreviations**

**C4ISTAR** – Command, Control, Communications, Computers (Computing), Intelligence, Surveillance, Target Acquisition and Reconnaissance

**UAV** – Unmanned Aerial Vehicle

**KRC -** "Killer Robots Campaign" - the Campaign Against Killer Robots

**List of Figures**

# Chapter One

## Introduction

In modern warfare efficacy of a military organization is dependent upon its C4ISTAR (Command, Control, Computing, Communications, Intelligence, Surveillance, Target Acquisition and Reconnaissance) methodologies, infrastructure and practice. A premium is placed on the ability of a military/security organization, to obtain empirically validated ( e. g. Battle damage assessments) tactical objectives with the least force, least cost and within stipulated time windows. The factor of least cost and proportionality of force in military operations, may be achieved by way of deploying autonomous war-fighting agents such as robots, in some operational instances.

Deployment of lethal autonomous weapons allows for greater risk exposure, at greatly minimized costs in terms of resources such as lives of mortal irreplaceable human operatives, there is also more efficient automated execution of military commands within reduced time-frames and with greatly reduced costs to humans by way of less mental stress, less financial strains and lower psychological burdens. But the emerging benefits of war-fighting automation has got its challenges, these being:

1. establishment of appropriate architectural frameworks to ensure that automation in war-fighting is conducted by hardware platforms and software environments, that are of excellent operational characteristics and design qualities
2. Production of automated war-fighting agents that are not only technically efficient, but also effective in operating within the bounds of local legislation and international humanitarian law.

### 1.1 Statement of the Problem

In today's conflicts it is ever more difficult to distinguish whether military force or police-based law enforcement suffices to neutralize an adversary, especially of an asymmetric-type. Given that the area of operations in an asymmetric conflict may be expansive and the actors sparsely situated within the battle-space, there is a need to utilize "smart" methods that embrace network-integrated sensors, social network visualization, human intelligence analysis, computers, actuators and systems, to increase the efficacy in tactical operations. A commander not only seeks to conduct operations, but also needs to ensure detailed data outflows are obtained from tactical operations to aid future operations and/or training sessions e.g. by way of deploying solutions based on Bayesian Modeling

Deployment of autonomous war-fighting agents whose functions are based upon computing information systems is likely to meet and exceed expectations of a command, but only if these vehicles are designed within the appropriate Architectural Frameworks. The multifaceted nature of conflict today in land, air and naval environments demands that automation in combat aggregates disparate gains to obtain an overall military objective.

### 1.2 Research Question

The general question is: What would be a generic architectural framework for designing, developing and implementing autonomous war-fighting agents for aggregation of gains obtained via multifaceted tactical environments, be? The practicality of the same would have to be demonstrated by way of analysis. Robust and versatile functionality would be critical in the realm of military/security operations.

### 1.3 Objectives

The research project arising from this study, establishes a C4ISTAR Architectural Framework for designing, developing and implementing autonomous tactical/war-fighting agents that are network-integrated. The autonomous tactical agents developed within the Framework shall have appropriate :

1. Systems Architecture Frameworks
2. Operational Architecture Frameworks
3. Technical Architecture Frameworks
4. Appropriate theoretical demonstrator modules

### 1.4 Justification

The research work shall establish and layout vital concepts that ensure that its generic autonomous war-fighting agents are safe to use and effective in executing any prescribed tactical operation. The originator of the study is cognizant of the fact that Kenya, the country within which the research is to be done is considered to be a third world and underdeveloped nation.

The architectural framework to be prepared shall have a stimulating effect in creating well engineered systems and products such as : unmanned aerial vehicles, missiles, advanced radar systems, jet-engines, electronic computing systems, robots, sensor-technologies, and many other potentially low-cost investment but high return systems. When these components are network-integrated there shall create multifaceted combat technologies for ground, air and naval operations that have appropriate Machine Learning algorithms and Tactical Heuristics.

# Chapter Two

# Literature Review

An architectural framework for a military type automated system has three primary components, namely:

Technical Architecture

Operational Architecture

Systems Architecture

These specify the design, interlinking, functionality and usage of technologies such as: Actuators, Radars, Sensors, Circuits, Data-links, Image/schematics matching, Launchers, Platforms, Projectiles/Missiles, Navigation Systems, Computer Programs; the objectives within the confines of this proposed Research Study would be to design and develop concepts for operating of autonomous war-fighting agents. To demonstrate the robust grounding of this proposed research work, Journal Articles, White Papers, Books, Primary Industrial/Institutional sources, have been reviewed and analyzed, a sample of which is rendered in this study.

## 2.1 A Sample of the Data-set

Automated control systems are dependent upon perceiving their operational environments with sensors, executing digital signal processing of sensor data streams, analyzing the deduced "perceptions" with preloaded heuristics and generating resultant control decisions which are used to manipulate their operational environments by way of actuators or additional projected autonomous agents. Upon actuation of a control decision, the sensors obtain feedback of extent of compliance or non-compliance thereby creating a feedback loop. Heuristics, past perceptions, recent perceptions, executed control decisions, and other system parameters are stored and manipulated in the memory of the automated control system.

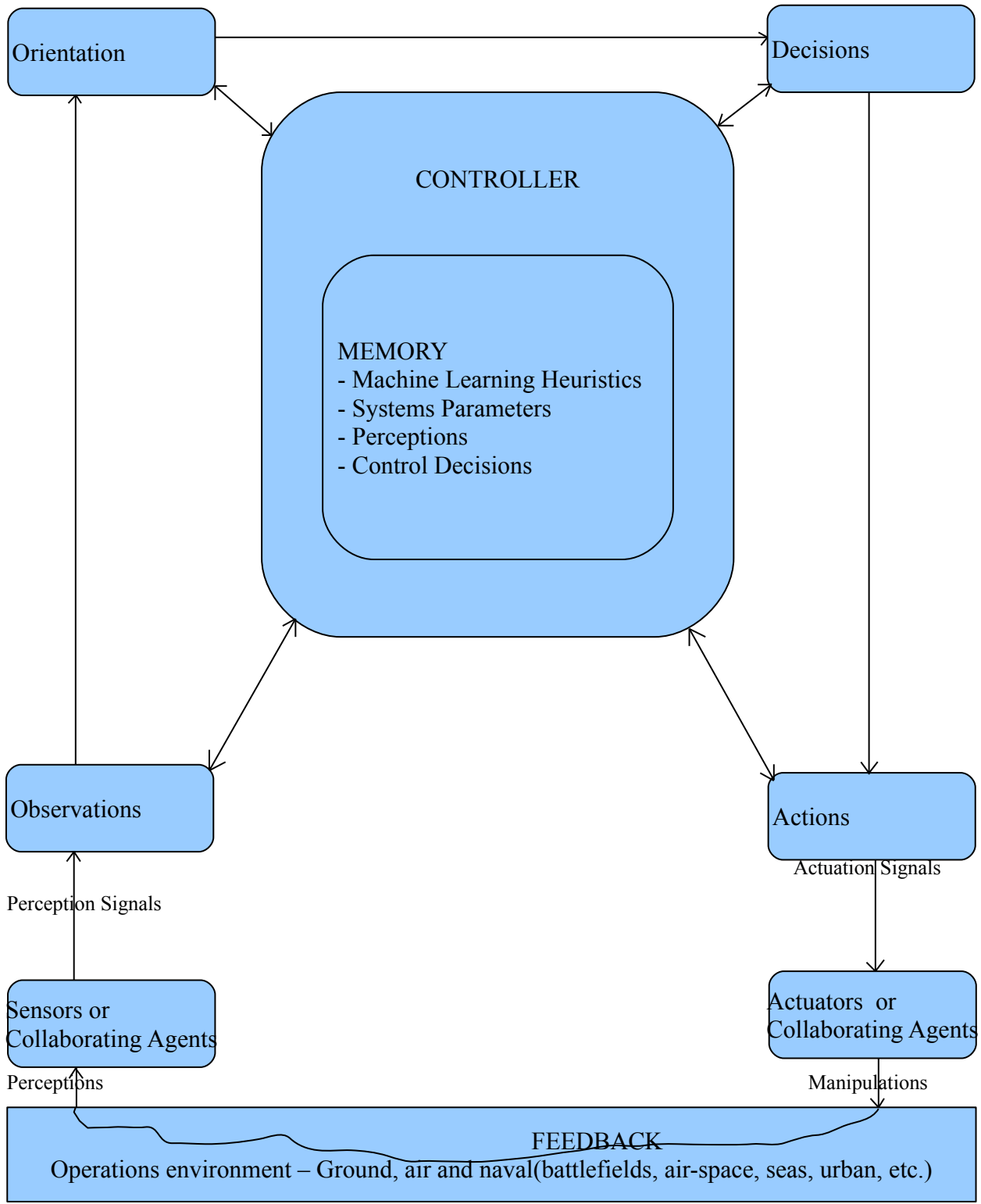

Figure A: Automated war-fighting/Tactical Agent Schematics

The concept automation is ably illustrated by Distefano *et al* [1] and Grant and Kooter [2]. Rizwan [3] elaborates how the Revolution in Military Affairs, has created an enabling environment that can today be associated with network-integrated and autonomous war-fighting capabilities. This is highlighted by the way in which the microchip has come to be a critical component of very many military systems. Rizwan goes on to detail the "Joint Force 2010" concept as postulated by General Shalikashral. Quantum computing and its related nano-components also offer new avenues for innovation in the realm of autonomous war-fighting

Critical to the "Joint Force 2010" concept is the theme of Total Dominance i.e. Dominant Manoeuvre, Precision Engagement, Full Dimension Protection and Focused Logistics, these have also come to be the underlying concepts in Architectures of Autonomous Weapons.

In the realm of Unmanned Aerial Vehicles (UAVs) there is an ever growing embodiment of autonomous war-fighting Dominant Manoeuvre is exemplified by the way in which disparately fielded UAVs can be automatically focused to obtain one particular military objective. Precision Engagement is today demonstrated by interdicting targets with use of "smart"/autonomous munitions, while Full Dimensional Protection is illustrated by broad ranging integrated systems engineering practice such as materials selection, electromagnetic hardening, electronic counter-measures, etc. Focused Logistics demands precision timing, navigation, essential payload and optimal workload.

Efficiency in war-fighting has a deterrent effect on potential adversaries but pursuit of Total Dominance via the Autonomous war-fighting track has got its firm opponents. KRC ("Killer Robots Campaign") Statement [4] lists some of the most vehement opponents of autonomous robots in warfare. But warfare of the autonomous type is not very easily defined, the "split hair" arguments are the basis of highly discriminative application. To the common man an autonomous robot moves around continuously searching for targets, after which it automatically qualifies and interdicts the target. One would imagine a robot moving and shooting at individuals. But even a passive homing anti-tank missile without control from the launcher is an autonomous robot once it is fired at a target – it is a robot which has a course of action(s) based on preloaded heuristics, and so are "smart" artillery munitions or even seismic sensing detonated munitions.

From a lethal autonomous weapons proponent's perspective several technological methods may be deployed to address humanitarian concerns, some of these are:
1. continuous updating of robotic algorithms with new algorithms and knowledge via data-links
2. "Perfect algorithms" that can address any tactical scenario and contingency
3. Building "humanitarian algorithms" for aborting tactical operations, when unforeseen challenges arise
4. Utilizing several different sensor types to verify a target before action
5. Quality assurance standards and inspections for autonomous war-fighting systems design, development, implementation and operation. This in turn reduces technical failure rates and malfunctions. *Committee on Armed Service United States Senate [5] Report* addresses the issue of standards and quality assurance especially in the realm of electronic components, circuitry and related systems. The proliferation of counterfeit electronic components in the global supply chain, has impacted negatively on the systems business by way of downtime increase, malfunctions, technical failures and related catastrophic disasters. Military forces such as the Armed Forces of the United States of America, have become proactive at identifying the problem of electronic component counterfeits within their systems and are demanding that manufacturers abide by certain standards.

Components that are integrated to form a system are vital, and equally vital are the people who deploy the technology and the processes/policies followed in deploying technologies. Processes include planning, design, development, implementation, training, operation, maintenance and upgrades. Personnel must have the requisite skills as acquired by way of training, education and experience. The triad of People, Processes and Technology is captured in Fawcett [6] where the researcher states that Russian war planners concluded that in the Gulf War (Operation Desert Storm) against Iraq, the Allies prevailed not only because they had the requisite technologies but more so that they were well trained (equipped with the relevant skill-sets) in addition to the hi-tech weapon systems. The paper alludes to the issue of motivation/willingness as the Allied Forces comprised of volunteer professional soldiers while those on the Iraqi side were mainly conscripts, this further captures the issue of combatant credibility.

We may extend personnel/people issues to those of honesty or otherwise – what is the credibility of the persons taking part in an operation? Credibility is critical because in the realm of fully lethal autonomous weapons, there are no soldiers or operators, they may only serve in roles as programmers, analysts, and designers who pre-load the autonomous weapon system with the relevant programming algorithms and logistics. Vertegaal [7] renders a system analysis and design concept developed in an era where the human user, was the "termination" and "initiating" point in military tactical networks. This concept also holds true for Hon [8], with the critical difference being that the prosecution of multifaceted warfare in a network-integrated manner was to be conducted by operatives linked to knowledge-based elements. Though not explicitly mentioned therein, it is not difficult to conceive a network in which those elements are lethal autonomous weapons or their supporting systems.

With growing demand for network-integrated tactical systems, the integrity of data-links is an issue, for example:
1. Credibility of equipment/services suppliers
2. Technologies availed by suppliers
3. Standards set by government bodies

Ruppersberger [9] also touches on the dimension of supplier credibility as is the report [5] of the United States Senate Armed Services Committee, analyzing the infiltration of counterfeit electronic components into the U.S. Military Supply Chain, there was special scrutiny given to suppliers from sources outside the United States of America. In this report [9] it was the Telecommunication Service Providers who came in for special scrutiny, the likelihood of espionage and sabotage via their networks and related equipment is evaluated. Systems and technical architectural frameworks must address supply chain risks if operational risks are to be minimized.

Over open networks such as the Internet, or any other potentially open communications media and protocols, the likelihood of adversaries seeking to subvert the systems functions arises. The downing of a United States Air Force drone over Iran is a demonstration of this norm [10]. These challenges are likely to come in two different forms, that may be combined in some instances, they are:

1. Use of malware to subvert an autonomous system
2. Interactive interventions by a human person via a network which is connected via electromagnetic waves

The first type of these is detailed by Sabovich and Borst [11], that advocates for the deployment of various systems engineering techniques, in preventing or containing malware challenges. Research arising from this study shall also cover a number of approaches in containing the threat of interactive intrusions. Emphasis shall be placed on the development and justification of new policy types as the primary basis of shielding networks hosting autonomous war-fighting agents from unauthorized external interactions or attacks.

Gompert and Saunders [12] explores issues of cyber-warfare in the context of a potential conflict between the Peoples Republic of China and the United States of America. This is a theater of the vaunted United States Military Strategy that is Air-Sea Battle, a concept dependent on processes such as automation and digitization. It further highlights the issue of network security in the realm of autonomous war-fighting Other additional concepts that are explored as those of mutual deterrence in the realm of cyber-warfare.

The concept of allowing for the redirection and reprogramming of autonomous lethal weapons that are fielded is important. Even with the inherent risks of hacking and malware, there are benefits in "flipping the script" and "shifting the goal posts" via a data-link when the "game" that is war is well underway. Military operations in most respects are dynamic, potentially unpredictable and require prompt decision making to cope with emerging threats and opportunities. It is very difficult if not impossible to come up with systems heuristics that can handle all manner of potential contingencies. Even human beings when fielded in a tactical operation often engage in collaborative thinking/group work or consultations via radio/data-links, when unforeseen contingencies are encountered.

Data from lethal autonomous weapons would be streamed back to a commander at a center, where the original intentions and objectives of unleashing such a system into the open environment that is a battle-space are known. The consoles at command stations are continuously monitored with a view to determining if strategic and tactical military objectives are being obtained. Blanco [13] offers empirically validated technical solutions, for determining whether those monitoring consoles are alert and aware of the overall situation. There is no limit to potential application of the same technologies at command centers where staff are monitoring feedback from autonomous war-fighting systems in the field. The alertness/awareness of a commander is also vital in determining if there is an ongoing or successful attempt, at disabling or disrupting the original preloaded objectives of an autonomous weapon by a hacker or malware.

## 2.2 Conclusion

Traditional military rules and regulations have been formed for well over a century. They are generalized as International Humanitarian Law, that is manifested by the Hague and Geneva Conventions, etc. it is not easy to get a change of these statutes. The widespread availability and affordability of dual use robot technology components, and the prevalence of robotics in almost every human endeavour makes it highly unlikely that an effective ban on lethal autonomous weapons shall ever come into effect as an integral part of what is today known as International Humanitarian Law. A prudent military force must be prepared for the eventuality that is autonomous war-fighting The experience of this researcher in this the domain of training military/security forces, is based not only on hind sight but also with the credible foresight that the Kenya Defence Forces and its Allies should not lag behind the world's fast changing military operations theatre that is embracing autonomy in tactical operations.

Implementing Lethal Autonomy with the current technological limitations implies that systems are not as highly discriminative when picking and engaging targets as may be desired by a commander. They may require the input of humans "in the loop" to make a final decision in some cases, while other systems like fully automated anti-ballistic missile defense system may not need human input though they do offer the option. Military weapons and/or their uses must meet certain critical International Humanitarian Law criteria:

1. They must be capable of making a distinction whether a potential target is a friend, neutral or a foe
2. All military attacks must be of a justifiable nature, e.g. wholly unprovoked attacks are considered to be war crimes, another element of law is that weapons used must have only sufficient effects, over-kills are illegal.
3. Battle injuries caused by weapons must not be aggravated beyond the extent of neutralizing a combatant, e.g. Weapons must not cause intentional infections or intentionally untreatable injuries
4. Weapons must allow only targeted and highly discriminative use
5. Adequate care must be taken not to kill or wound non-combatants, that includes giving them warnings
6. Weapons that cannot have limited usage must not be deployed

Robots are common feature in human lives for a wide range of functions/applications e.g. transport, medicine, etc. Can it be claimed that there should be a pact between human and machines? Or are machines an extension and manifestation of human intelligence? Lethal autonomy increases the ease with which human life can be taken or degraded, the decision to deploy such technology must not be taken lightly. But there are still a wide range of factors e.g. experience and intelligence borne by human war-fighters that cannot be easily codified for use in the domain of warfare. Nevertheless the ease with which electronic components such as actuators and chips can be obtained and programmed implies that lethal autonomy is at least inevitable in the domain of non-state actors, it offers them an anonymous means of projecting and executing their power unlike any other time in history.

Apart from surveillance and target acquisition, robots may need to analyze the intentions of human beings. This is one of the greatest challenges, other challenges include codification of empathy with human persons. It is only of late that technology for remote credibility analysis of humans are being developed and deployed e.g. stress analysis via thermal imaging but these are not yet robust to the extent of being used with consistently good results. Contrary to popular belief there are a wide range of fully autonomous weapon systems that are currently in use by military forces e.g. Sensors that detonate explosive charges when they are interfered with, but of the current fielded systems there is none that can readily tap into the human mental process and know how to harness its feedback.

A critical point of interest for most military and security forces in the domain of war-fighting would be safety and reliability issues, autonomous robots remove the human user risk. But the variables are still very many, and it is too difficult to determine which technologies shall result from the current emerging experiences in the course of tactical operations. Human beings are capable of harnessing fine grained ethical and legal decision making processes that are difficult to technically replicate, this is the major challenge to developing lethal autonomy, there are operational circumstances that are difficult to predict/foresee and codify until they actually occur. The potential broad spectrum applicability of lethal autonomy shall disadvantage any military/security force that has its State ratify any "global" ban. Arguments about the benefits and paybacks of deploying or not deploying autonomy are becoming mainstay in evermore domains of human life, for knowledgeable non-state belligerents lethal autonomy is an option.

Currently the United States of America's Department of Defense was one of the first institutions to come up with a directive policy [14] for autonomous weapon systems usage. It answers the question that is the yearning of many i.e. A human in the loop at the point of firing a weapon on an autonomous platform, but as robots become ever more fast in decision making, this cardinal principle shall inevitably be compromised with a view to interdicting threatening targets before they can do harm to interests protected by lethal autonomous weapons. There shall be a much faster target acquisition and interdiction loop, may be in the scale of milliseconds and human contribution to the same shall be minimal if not impossible.

Chapter Three

Research Methodology

Lethal autonomous weapons are spinoffs of technical advances in fields such as robotics, computing, navigation, etc. Versatility, flexibility, ingenuity and urgency of deployment of newly emerging technologies inevitably applies pressure on weapons systems engineers to come up with appropriate systems integration/engineering concepts that exploit advances in the domain of sci-tech. This inter-disciplinary approach to providing autonomous agents has a lot of traction with commanders who can get ever greater tactical payoffs with ever diminishing human risk exposure. Searches for information and data that actualizes the benefits of lethal autonomy takes place over diverse specialties and various experts may not be forewarned of the use of their research in this domain.

To come up with a comprehensive and broad spectrum Architectural Framework for lethal autonomous weapons, a lot of white papers, journals, books, theses, and other research materials shall be reviewed repeatedly by the researcher.

### 3.1 Qualitative analysis of the Research Problem

The objective of the proposed research work develop a C4ISTAR Architectural Framework that guides in the development of weapons that manifest lethal autonomy. There is already an ever growing body of literature on lethal autonomy, that is broad and rich in insights. The method adopted for the study is the Qualitative Analysis. The specific approach shall be the Grounded Theory Method of Qualitative Analysis.

### 3.2 Research Methodology – Grounded Theory Method

Research work undertaken by way of the Grounded Theory Method shall be done as follows:

1. Conducting a relevant literature search from libraries, databases, digital libraries, media monitoring, etc. to produce an initial examinable data set. Consultations with various experts shall be undertaken during the whole course of my research work.
2. This initial data-set shall be reviewed by the researcher for relevance and potential utility, with a view of scoring the sources for relevance and making a determination as to the sufficiency of the data-set.
3. The qualified data-set shall then be obtained and stored in hard or soft copy format and cataloged/indexed by the researcher for easy/quick reference.
4. This qualified data-set shall be reviewed severally with a view to enabling the researcher to understand its contents and contexts, and to draw-up initial working summaries to be used in the qualitative analysis [17].
5. More sources shall be sought to support or disprove the initial working summaries.
6. An initial research paper is then drawn up by way of qualitative analysis of the refined data-set is the Grounded Theory [20] developed by way of appreciating and integrating the contexts and contents of the disparate research sources. In line with the principles of Leshem [21] a conceptual framework is drawn-up based upon the research work, that details the interactions between the variables discovered in the course of research.
7. The initial research paper is then reviewed by selected Academic Readers, to determine the extent to which the initial research study is manifested via the research work and if the work is adequate in terms of form, contexts, contents and knowledge.
8. Readers give the researcher feedback which is incorporated into a review of the paper after which the Paper upon approval is submitted for publication

### 3.3 Dissemination Strategy

The research work shall be published in several papers, some of these shall have direct peer-review, and others published on expert refereed online platforms or online platforms that allow for peer citation counts. The major by-product of the research shall be robust lethal autonomous weapons systems, these shall be conceptualized, designed and developed by outfits selected by the researcher and used in accordance with existing laws. Academic conferences and seminars shall also offer the researcher a mode of peer-reviewed dissemination. Academic work such as training and education shall also provide a powerful method for the research dissemination.

### 3.4 Conclusion

There was a wide range of potential research resources, this made it feasible for the research work to be undertaken. Some may view the Grounded Theory Method of conducting research as a subversion [18] of quantitative methods, but that is not the case as the research data-sets obtained qualitatively also contain quantitative and empirically generated data-sets that can only be analyzed by way of their original quantitative analysis mechanisms within a Grounded Theory Method. Research is basically a learning process [19], and in all likelihood new approaches and knowledge that was not envisioned in the initial objectives may be obtained.

### 3.5 Biographies

1. Reader:: David K. Ngondi, works at the Directorate of Police Reforms of the Kenya Police Service and has several decades of experience in a broad-spectrum of security and para-military operations. He posses a Master of Arts degree in Security Studies from the University of Hull in the UK.

**IMPORTANT NOTE: Times New Roman, font size 10 and Harvard referencing [15][16]used on this document.**